**Title:** TRACED-X enables in vivo imaging of cellular microstructure, intrinsic extracellular diffusivity, tortuosity, and membrane permeability in human glioma patients

**Authors:** Joshua K. Marchant[1,2], Hong-Hsi Lee[1,3], Elizabeth R. Gerstner[1,4,5], Susie Y. Huang[1,2,3], Bruce R. Rosen[1,2,3]

1. Athinoula A. Martinos Center for Biomedical Imaging, Department of Radiology, Massachusetts General Hospital, Boston, United States of America
2. Harvard-MIT Program in Health Sciences and Technology, Massachusetts Institute of Technology, Cambridge, United States of America
3. Harvard Medical School, Boston, United States of America
4. Department of Neurology, Massachusetts General Hospital, Boston, United States of America
5. Mass General Brigham Cancer Institute, Boston, United States of America

Corresponding author: Joshua Marchant, jmarchan@mit.edu

Preprint under review for *Magnetic Resonance in Medicine*.

## Introduction

Accurate characterization of tumor microstructure remains a central challenge in neurooncology. While diffusion MRI-based estimates of cell size and density can aid glioma grading and treatment planning,[1–4] the properties of the tumor extracellular space (ECS) which govern transport, infiltration, and treatment response remain poorly characterized in vivo. Noninvasive characterization of the tumor ECS could provide critical insights into drug delivery (including theranostics) and interstitial transport[5–9] and identify changes in the tissue microenvironment preceding gross tumor cell infiltration.[10] Water exchange between intracellular and extracellular compartments further modulates the diffusion MRI signal, and, if adequately modeled, could provide complimentary functional information regarding altered membrane permeability in tumor regions. A diffusion MRI model which provides robust characterization of cellular microstructure, ECS diffusivity, and water exchange could open new doors for characterizing and targeting the glioma microenvironment in vivo.

Quantitative diffusion metrics have already shown remarkable progress in revealing tumor microstructure,[11–19] identifying actively growing or necrotic regions[18,20] and informing diagnosis and treatment planning.[6,7,20] Biophysical models previously developed for inferring tumor microstructure from diffusion MRI data include VERDICT,[17] IMPULSED,[11,13,14,21–24] POMACE,[25] MR-Cytometry,[15,26] and CEXI.[27] Implementations of the basic two-compartment (intracellular and extracellular) diffusion model have demonstrated efficacy in estimating tumor cell density (cellularity) and cell size.[11,12,28] For example, Jian et al have demonstrated the utility of these parameter estimates to predict early microstructural treatment effects in tumor models.[23] Grussu et al demonstrated clinically-feasible measurements of liver tumor cell size and cell density with variations of the two-compartment model.[28] Cell size measurements benefit from multi-b-shell, multi-diffusion time/frequency diffusion protocols, which provide sensitivity to a greater range of intracellular diffusion correlation times.[22,29]

Despite these advances, previous models have some limitations. Most assume a uniform cell size without consideration of size distributions. The exchange-focused CEXI model demonstrated bias in parameter estimates due to variable radius distributions.[27] A notable exception is the aforementioned MR-Cytometry method, which estimates a non-parametric distribution of tumor cell radii using a discretized fitting formalism and the combination of oscillating gradient and pulsed gradient spin echo acquisitions.[15] While the model performed well in its primary goal of measuring cell size distributions in the context of various tissue types, the authors report that MR-Cytometry ECS diffusivity was poorly constrained in its description of time dependence and in regimes with low water restriction (e.g., necrosis).

An additional limitation of existing approaches lies in the treatment of the extracellular compartment. Incorporating extracellular diffusion time dependence and non-Gaussian signal decay into the signal model remains a challenge for characterizing both healthy and malignant tissues. Even for simple micro-geometries like packed spheres, no analytical solution is known to describe extracellular diffusivity time dependence at intermediate diffusion times (~20-50 ms) on clinical scanners.[25,30] Some models ignore extracellular diffusivity diffusion time dependence altogether, while others employ approximate solutions, suitable only for very short-[25,31,32] or long-time regimes.[25,30,33] Furthermore, kurtosis-based approaches acknowledge the nonzero effect of higher-order ECS diffusivity terms in the diffusion signal decay, but most compartmental models neglect higher-order terms to avoid additional free parameters. Thus, models may bias other free parameters to account for the unmodeled non-Gaussian extracellular diffusivity.

Given the complexity in modeling time-dependence and higher-order extracellular diffusivity terms, most models to date focus primarily on accurately estimating the cell radius and volume fraction; extracellular diffusivity remains a relative or ill-defined metric. This limitation is particularly amplified in the study of drug diffusivity within tumors, where the ECS plays a critical role in therapeutic efficacy. Reliably quantifying extracellular *tortuosity* (a geometric hindrance factor which reflects path elongation due to boundaries[8,34]) and the *diffusivity of the ECS* could shed light on tumor regions where drug penetration is likely to be reduced or where subpopulations of densely packed tumor cells may evade therapy.[35]

Finally, few compartmental tumor microstructure models to date incorporate water exchange. Broader glioma literature illustrates that aquaporin expression and associated transmembrane water exchange rates vary across glioma types, grades and intratumoral regions, providing a biological basis for heterogenous membrane permeability and suggesting its value as a noninvasive biomarker.[36–39] Furthermore, neglecting the contribution of exchange may bias estimates of other parameters.[27] The early FEXI model recovered an apparent exchange rate in glioma tissue but without identification of tumor microstructure.[40] Emerging work using free gradient waveforms deemed ResEx showed the heterogenous organization of estimated water exchange time in glioma patients.[41] Other prominent compartment-based tumor models, such as VERDICT[17] and IMPULSED,[12] neglect the effects of water exchange. The CEXI model[27] incorporated exchange using an approach based on the Kärger model. Previous work notes that, while the Kärger model is appropriate in some regimes, its assumptions break down in multiple conditions relevant to tumor imaging with advanced gradient systems – at sufficiently short diffusion times and for cells of relatively large radius or high permeability.[27,42] As suggested by previous white matter-based models,[43] a Monte Carlo-trained representation of membrane permeability has the potential to overcome these limitations and produce more stable estimates of water exchange.

To address these challenges, we introduce TRACED-X (**T**ortuosity and **R**adius **A**ssessment via intra**C**ellular and **E**xtracellular **D**iffusion with E**X**change), a biophysical model trained using Monte Carlo diffusion signals which incorporates heterogenous cell populations, time-dependent diffusivity, and cell membrane permeability. TRACED-X enables the estimation of cell volume fraction, the cell size distribution, intrinsic ECS diffusivity, cell membrane permeability, and by extension, the ECS tortuosity. Here, the "intrinsic" ECS diffusivity intends to describe the true extracellular diffusivity without the impact of modeled cell boundaries (i.e., solely extracellular matrix (ECM) restriction) or diffusion time effects. Tortuosity can then be derived as the ratio of intrinsic ECS diffusivity to effective diffusivity at long time-scales. TRACED-X builds upon the initially proposed TRACED formalism[44,45] with the inclusion of exchange. We are unaware of any previous formalism which enables the estimation of intrinsic ECS diffusivity and tortuosity in addition to cellular microstructure and membrane permeability in tumor microstructure.

We first present the Monte Carlo simulations and neural networks underlying the TRACED-X model and parameter fitting. Simulation-based benchmarking simulations alongside multiple comparator models are described. Then, we apply the TRACED-X to multi-b-shell, multi-diffusion time data acquired on the Connectome MRI scanner with maximum gradient strength of 300 mT/m in test set of 8 patients with gliomas of various grades, demonstrating its feasibility and potential to reveal novel contrasts related to tumor interstitial structure and exchange.

## Methods

### *Model Formulation and Parameter Fitting*

In the TRACED-X formalism, voxels are simulated using randomly-packed semi-permeable spheres with variable cell size and volume fraction. To generate a signal dictionary for learning the parameter inversion, we employed Monte Carlo-based PGSE diffusion simulations building upon previous code[46,47] in spheres with radii $R_n$ sampled from a lognormal distribution (mean $\bar{r}$, radius coefficient of variation CV). The dictionary was initiated with 5,376 scrambled Sobol-sampled geometries, each with $N = 999$ spheres, within the following parameter ranges: $f_c \in [0.02, 0.63]$, $\bar{r} \in [0.5, 20]$ μm, $CV \in [0, 0.5]$, $D_{ec} \in [0.05, 3]$ $\mu m^2/ms$. A lognormal distribution was chosen as a reasonable approximation given previous literature showing lognormally-distributed cell size patterns across cell types.[48–52] Only geometries with $\bar{r} \leq 14$ μm were used in the final training set. Each geometry was simulated with a baseline zero-permeability ($\kappa = 0$) case and 15 nonzero membrane permeability values. Each geometry sampled one randomly selected nonzero $\kappa$ from 15 equal-width bins in $\log_{10} \kappa$ space, where $\kappa \in [10^{-4}, 10^{-1}]$. The intrinsic intracellular diffusivity was fixed at $D_{ic} = 1$ μm$^2$/ms to reduce the number of free parameters, consistent with previous strategies.[53–55]

Monte Carlo diffusion simulations were performed using the b-values, gradient pulse width ($\delta = 8$ ms), and two diffusion times ($\Delta = 19, 49$ ms) that matched our in vivo data acquisition (described below). We utilized 50k random walkers and 5E-4 ms-time steps following signal convergence tests, which demonstrated signal accuracy levels at or below the noise floor. Normalized signals were rejected if any shell value was zero, negative, or exceeded 1.05 (~9% rejection rate). Total time for the 117, 664 runs on Quadro RTX6000 partitions was approximately 324 GPU-hours (approximately 10 hours of ideal run time using 32 simultaneous GPUs). The final dictionary was divided into approximately 80% training, 10% validation, and 10% test geometries.

Separate MLP-based networks (3 hidden layers, 160 neurons/layer) were trained in PyTorch[56] to perform supervised amortized inversion of the simulation-derived signal dictionary for each TRACED-X parameter $\boldsymbol{\theta} \in \{f_{cell}, \bar{r}, CV, D_{ec}, \kappa\}$ in the spirit of previous dictionary-based training approaches.[43,57–59] Details regarding the training strategy and patient-informed Rician noise in the simulated training dictionaries are outlined in the Supplementary Information.

The intrinsic extracellular diffusivity $D_{ec}$ represents the free diffusion coefficient of the ECM only, disregarding the impact of cell boundaries. The difference between $D_{ec}$ and free water diffusivity $D_{free} \approx 3$ μm$^2$/ms at body temperature therefore reflects the additional hindrance in the ECS coarse-grained[60] by diffusion, rather than the hinderance caused by the explicitly modeled cell boundaries. No known analytical model exists to describe the relationship between the *effective* diffusivity $D_{eff}$, tissue parameters ($D_{ec}$, $\mu_r$, $\sigma_r$, $f_c$), and the diffusion time Δ in the intermediate time regimes. Quantification of $D_{ec}$ and $D_{eff}(\Delta)$ permits the description of extracellular tortuosity, which we define as:

$$\tau = \frac{D_{ec}}{D_{eff}(\Delta \to \infty)} \quad . \quad [1]$$

Therefore, to bridge the gap between $D_{ec}$ and $D_{free}$, we performed additional Monte Carlo simulations of extracellular diffusion only using the aforementioned 5,376 geometry dictionary and $\kappa = 0\ \mu\text{m/ms}$. This dictionary was designed to train a small neural network (2 hidden layers, 10 neurons/layer) to learn the relationship between the input parameters ($D_{ec}, \mu_r, \sigma_r, f_c, \Delta$) and $D_{eff}$, as $D_{eff} = NN(D_{ec}, \mu_r, \sigma_r, f_c, \Delta)$. Diffusion times were uniformly sampled from 10 to 200 ms for each simulation, and a broad range of b-values were used ($\text{b} = 0, 250, 500, 800, 1200, 1600, 2000, 3500\ \text{s/mm}^2$) for robust fitting. In each simulation, $D_{eff}$ was calculated as the negative linear coefficient of a fifth-degree cumulant expansion of powder-averaged extracellular signal.[61] A reduced walker count and step size (12.5k and 8E-4 ms, respectively) were selected after convergence testing (< 1% error in derived $D_{eff}$ compared against 100k walkers/2E-4 ms step size). A variety of example microstructural conditions were simulated to investigate the behavior of $D_{eff}$ at different diffusion times. Because the neural network learns the diffusion time-dependence of $D_{eff}$, it uniquely permits the estimation of extracellular tortuosity due to diffusion hinderance between cells.

The direct estimation of the cell membrane permeability $\kappa$ also allowed for the derivation of the total exchange time[27,42], defined as

$$t_{ex} = \frac{1}{k_{tot}} = \frac{(1 - f_{cell})\, R_{32}}{3\kappa}\ , [2]$$

where the Sauter mean radius $R_{32}$ [27] for a lognormal distribution is found as:

$$\frac{\langle R^3 \rangle}{\langle R^2 \rangle} = \bar{r}(1 + CV^2)^2$$

*In vivo MRI*

Diffusion weighted, FLAIR, and pre-contrast T1-weighted images were acquired for eight mixed-grade glioma patients (5 males; 36±13 years old) on a 3T Connectome 1.0 MRI scanner (MAGNETOM Connectom, Siemens Healthcare, Erlangen, Germany[62]). This dataset was presented and analyzed using tensor analysis in a previous work.[63] Proof-of-concept fitting was performed for all 8 patients, whose pathologic diagnoses are presented in Table 1. The analyzed images were acquired before resection and chemotherapy/radiation treatment. In the case of Patient 1, images post-resection/pre-chemoradiation (Month 2), and post-chemoradiation/mid adjuvant temozolomide (Month 8) were also assessed. 12 b-values across two diffusion times were used ($\Delta = 19$ ms: $\mathrm{b} = 100, 500, 1300, 2600, 4200, 6000\ \mathrm{s/mm^2}$; $\Delta = 49$ ms: $\mathrm{b} = 700, 1400, 3800, 7300, 12000, 17800\ \mathrm{s/mm^2}$), with a pulse width of 8 ms, 32 and 64 directions per b-shell for b-values < 2400 s/mm$^2$ and ≥ 2400 s/mm$^2$, and 2-mm isotropic resolution (see the aforementioned prior work[63] for full acquisition details). Total scan time per patient including was approximately 56 minutes. Noise maps were estimated using the MP-PCA denoising algorithm (via MRtrix3[64,65]). Anatomical FLAIR, pre/post contrast T1-weighted MRI, apparent diffusion coefficient (ADC) and direction-encoded color fractional anisotropy (cFA) maps are also displayed. Acquisition and derivation details are described in the Supplementary Information. The FLAIR hyperintense ROI was manually defined by a neurooncologist. For Patient 5, normalized subtraction of same-session pre- vs. post-contrast T1w MRI was used to identify the contrast-enhancing regions,[66] followed by manual exclusion of vessels using 3D Slicer (http://www.slicer.org[67]).

*Simulation Benchmarking*

Simulation-based benchmarking of TRACED-X and three comparator models was performed using held-out Monte Carlo simulations. First, a basic two-compartment model denoted as "Simple" without consideration of a cell radius distribution (i.e., $CV = 0$) as utilized in previous work[28] was considered:

$$\frac{S(b)}{S(0)} = f_{cell} \cdot S_{ic}(b, D_{ic}, \bar{r}) + \ (1 - f_{cell}) \cdot \exp\left(-b \ \cdot D_{eff}\right). [3]$$

The signal $S_{ic}$ is calculated based on the Gaussian-phase-approximation (GPA) solution.[68] Next, the CEXI model, which builds upon the 2-compartment framework while incorporating exchange between the intracellular and extracellular compartments using a Kärger-based strategy.[27] Lastly, we developed a variant denoted CEXI-CV, which incorporates a lognormal-radius distribution into the intracellular portion of the forward model to enable a fair comparison against the lognormal radius-distribution-trained TRACED-X model. The details of this formulation are described in the Supplementary Information. We note that only TRACED-X is capable of providing $D_{ec}$, while the other models report an effective or apparent ECS diffusivity. Each model was trained and evaluated using the patient noise-informed MLP-based supervised learning framework noted above.

To explore performance in different stress regimes, held-out MC-simulated challenge data were generated using conditional sweeps over radius CV, permeability, and ellipsoid aspect ratio $AR$ (ratio of polar to equator length) levels. The radius CV and $\kappa$ sweeps were conducted using 1,280 geometries, with Sobol-sampled distributions over the following parameter ranges: $f_{cell} = [0.02, 0.60\ ], \bar{r} = [4, 12]\ \mu\text{m}$, and $D_{ec} = [0.1, 2.95]\ \mu\text{m}^2/\text{ms}$. For the radius CV sweep, $\kappa$ was fixed at 0 $\mu m/ms$; for the $\kappa$ sweep, the radius CV was fixed at 0.2. For the anisotropy $AR$ tests, 200 packed ellipsoids were generated using a custom modification of the RMS sphere-packing method as described in the Supplementary Information.[46] A fixed grid of 10 $f_{cell}$ values spanning $[0.02, 0.60\ ]$, CV = [0.1,0.2,0.3, 0.5], and 5 alternate-seed packing replicates were used. Mean volume-equivalent radius and permeability were maintained at $\bar{r} = 5\ \mu\text{m}$ and $\kappa = 0.03\ \mu\text{m}/\text{ms}$. For all sweeps, patient-derived noise contamination was applied in the same manner as described above for noisy data training.

*K-Means Tumor Habitat Analysis*

To explore relevant voxel groupings, we conducted a proof-of-concept K-Means cluster analysis for 2 patients (Patient 1 and Patient 5) selected for their larger, spatially heterogenous lesions. Using the scikit-learn *KMeans* algorithm,[69] we conducted an initial diagnostic sweep assuming $k = 1, 2 \ldots 8$ clusters. Five independently seeded fits were performed with 20 candidate initializations each. We selected the highest $k$ resolution which also satisfied the following reproducibility criteria: global adjusted Rand index > 0.80 and worst-overall Dice score > 0.90 for each cluster and repeat initialization.

*Variable Acquisitions*

Finally, to explore translatability, we performed MLP regression training using an 8-shell, 2-Δ protocol (maximum b-value of 4200 s/mm$^2$, representative of an achievable acquisition on emerging FDA-approved high-performance gradient scanners, such as the Cima.X from Siemens Healthineers, Erlangen, Germany, or MAGNUS[70] from GE HealthCare, Waukesha, WI, USA) and a single-Δ (49 ms), 3-shell protocol (maximum b-value of 3800 s/mm$^2$; representative of achievable acquisition on commonly available high-end clinical scanners). This provides an approximation of model performance with different acquisitions but does not account for all scanner acquisition variations (such as TE, $\delta$, and Δ constraints).

## Results

*Model Fit Performance*

Test performance of the MLP regressors on held-out signals from the primary training data set showed strong performance for clean and noise-contaminated data (Figure 1a). (Note that inset statistics in Figure 1a include results over all $\kappa$ samples.) Noise-contaminated data was fit with reduced accuracy over varying permeability levels. Held-out $D_{ec}$ fits resulted in the highest fit accuracy of all parameters and showed minimal degradation with noise and permeability ($R^2 = 0.97, MAE = 0.1\ \mu m^2/ms$), while cell fraction and mean radius fitting showed mild degradation with noise over all permeability levels ($R^2 = 0.83, MAE = 0.05$ and $R^2 = 0.84, MAE = 1.0\ \mu m$, respectively). Radius CV was the most difficult microstructure parameter to recover ($R^2 = 0.3, MAE = 0.12$ with noise). Membrane permeability (fitted in log-transformed space as $\log_{10} \kappa$) fitting showed lower precision but consistent order-of-magnitude trend agreement ($R^2 = 0.50$, $MAE = 0.46$ with noise).

Using 1,024 held-out geometries with 8 diffusion times each spanning [10, 150] ms, the surrogate neural network for predicting $D_{eff}$ from $NN(D_{ec,}f_{cell}, \bar{r}, CV, \Delta)$ achieved $R^2 > 0.99$ and MAE = 0.006 $\mu m^2/ms$ (Figure 1b). The predicted $D_{eff}$ demonstrated variable time dependence in the short and intermediate Δ regimes (Figure 1c-d). As predicted by the diffusion coarse-graining effect,[33,60,71,72] $D_{eff}(\Delta)$ monotonically decreased with time and approached a plateau for $\Delta > 100$ ms for all three example conditions, suggesting the start of the long-time regime (where $\Delta \gg \bar{r}^2/D_{ec}$; indicated by $D_{eff}(\Delta \to \infty)$ in Equation 1). $D_{eff}$ increased with decreasing $f_c$ and increasing $\bar{r}$ and demonstrated increased nonlinearity for short Δ values.

TRACED-X fitting resulted in reduced median absolute error and interquartile range (IQR) compared to the comparator models for most parameter sweep combinations (Figure 2). The ECS diffusivity row demonstrates the model-specific ECS diffusivity ($D_{ec}$ for TRACED-X, $D_{eff}$ for other models) with respect to the *true* $D_{ec}$. While not a direct comparison, the results demonstrate the variability of model-recovered diffusivity compared to the true intrinsic ECS diffusivity. Over the swept radius CV levels, 2CM, CEXI, and CEXI-CV showed variable stability in fitted parameters. All Simple model parameter estimates had increased error with increasing CV and $\kappa$. Simple-fitted $f_{cell}$ and $\bar{r}$ showed competitive accuracy for low CV/$\kappa$. TRACED-X resulted in reduced median absolute error/IQR compared to CEXI and CEXI-CV when fitting $f_{cell}$ at high permeability and $\kappa$ over all permeability levels. Fitting the powder-average 32/64-direction signals for the fully-aligned ellipsoid geometries showed negligible bias for all models up to $AR = 1.6$.

*In vivo Patient Fitting*

TRACED-X maps were computed for the full brain volumes for all eight patients. FLAIR-estimated tumor volume are recorded in Table 1. Whole-brain results are demonstrated to prevent cropping of peritumoral tissue; however, we note that TRACED-X is not immediately applicable to healthy white or grey matter microstructures, and analysis is only performed inside the FLAIR-hyperintense ROI.

Single-slice maps of the fitted TRACED-X parameters for four patients are demonstrated in Figure 3, paired with FLAIR, T1w pre/post-contrast, ADC, and cFA maps. TRACED-X reveals quantifiable microstructural contrast and organization beyond the anatomical images. Patient 1 demonstrates a large, heterogenous Grade 3 astrocytoma lesion with distinct spatial regimes of elevated cell fraction, mean radius, and mildly elevated cell permeability, with diffusely elevated $D_{ec}$. Patients 3 (Grade 2 oligodendroglioma) and 7 (Grade 3 astrocytoma) show smaller, more homogenous lesions, with conspicuous low T1w /FLAIR spots coupled with low $f_{cell}$ and elevated $D_{ec}$, which may indicate focal cystic/microcystic, fluid-rich regions as previously described in both tumor types.[73,74]

Patient 5 (GBM) demonstrated marked heterogeneity in fitted microstructure, $D_{ec}$, and $\kappa$ inside the FLAIR ROI (Figures 3 and 4). The lesion exhibits a high-$f_{cell}$ core, elevated CV throughout the tumor, and layered $\kappa$ organization. Multiple map features in the tumor core demonstrate co-alignment with and extend beyond the CE region (Figure 4). The high $f_{cell}$ core (Zone A), followed by spreading regions of moderately elevated $f_{cell}$ (Zone B) and a lower $f_{cell}$ periphery (Zone C) is consistent with GBM microstructural organization and tumor cell invasion: the Ivy Glioblastoma Atlas project notes that GBM lesions typically exhibit a cell-dense tumor core (CT), infiltrating tumor (IT) with intermediate cell density, surrounded by the leading edge (LE) of the tumor where fewer tumor cells are present.[75] A similar gradient of increasing hyperintensity was observed in the FLAIR image from Zone A to C; thus, the reduced $f_{cell}$ in Zone C may be an indication that per-voxel volume fraction was increasingly dominated by edema near the tumor periphery. A reduction in fractional anisotropy (FA) was noted in the contrast-enhancing/CT-like region. Notably, the CT-like and LE-like regions of Patient 5's lesion showed decreased $\kappa$, whereas the intermediate Zone B had markedly elevated $\kappa$. This is also consistent with previous studies at the microscale: Civita et al. found that infiltrating GBM tumor cells in perineuronal satellitosis express more AQP1/AQP4 than tumor cells in the core (CT)**;**[76] Oishi et al. demonstrated that AQP1 expression increased GBM migration and invasion experimentally;[77] and De et al. identified AQP4 and GFAP were enriched in the IT compared to the peripheral LE region.[78]

Figure 5 illustrates the fitted $D_{ec}$, NN-generated $D_{eff}(\Delta \to \infty)$, and derived $\tau$ maps (Equation 1) for Patient 5. For all patients, extracellular tortuosity–cell fraction relationships ($\tau$ vs $f_{cell}$, using $\Delta =$ $150$ ms based on Figure 1) showed good agreement with EMT predictions for packed spheres [79–81] (Supplementary Figure S2 and S3), given by Archie's law

$$\tau = \alpha \cdot (1 - f_{cell})^{-\beta}\ , [6]$$

where EMT predicts $\alpha = 1$ and $\beta = 0.5$. Our Monte-Carlo simulations indicate excellent agreement with EMT predictions in the dilute regime ($f_c < 0.15$) (Supplementary Figure S2). For higher cell fractions, a reduced exponent of $\beta\ \approx 0.35 - 0.4$ was observed, deviating from the theoretical value of $0.5$. This discrepancy likely reflects the breakdown of effective medium theory assumptions at high packing fractions where interactions between densely packed spheres become significant. These results demonstrate the feasibility of a geometry type-agnostic strategy for combining $D_{ec}$ with neural network-generated $D_{eff}(\Delta \to \infty)$ to derive ECS tortuosity.

Whole-tumor mean, median, and IQR for all TRACED-X fitted and derived parameters in each patient are illustrated in Figure 6. Inside the FLAIR ROI, Patient 5 (GBM) showed elevated cell fraction, mean radius, radius CV, permeability, tortuosity, and reduced exchange time compared to other patients. Cell fraction, mean radius, CV, $D_{ec}$, and tortuosity were all elevated in the Patient 5 contrast-enhancing (CE) region compared to the FLAIR-minus-CE ROI, while permeability was moderately reduced.

The single-patient habitat analysis revealed reproducible voxel phenotypes for Patient 1 and Patient 5 with differing per-habitat parameter organization (Figure 7). The highest-partition stable solution criteria resulted in selection of k=4 partitions and k=5 partitions, respectively. Partitions or "habitats" are denoted as H1, H2 … HN. Remarkably, 91.2% of the contrast enhanced-ROI in Patient 5 was described by H5 and H6, with the higher-$\kappa$ H5 regime contributing most (74.6%). Lower cell fraction habitats of mixed permeability (H1 and H2) were enriched in the tumor periphery, while the central region of the lesion was predominantly described by H3-H5.

Longitudinal scans were assessed from Patient 1 (Figure 8). Twenty-seven days after partial resection, a transient region of T1w and FLAIR hyperintensity was colocalized with elevated $f_{cell}$ and $\kappa$ inside a larger zone of elevated $D_{ec}$. FLAIR and T1w-hyperintensity are hallmark features of hemoglobin conversion to paramagnetic methemoglobin in postsurgical hemorrhage, particularly in its mid stages. Previous work suggests that postsurgical hemorrhaged can develop or worsen as late as ~7.5 days post-operation,[82] and intratumor hemorrhaged hemoglobin degradation may occur more slowly in tumor tissues.[83] Blood clotting (dense concentrations of erythrocytes) is consistent with elevated $f_{cell}$. Previous work has independently measured human erythrocyte membrane water permeability at approximately $\kappa = 0.05 - 0.07\ \mu\mathrm{m/ms}$,[84] which aligns with our measured $\kappa$ values. Additional validation experiments will be required to corroborate these hypotheses, but these trends are consistent with the biophysical properties of tissues known to be present in this clinical scenario.

With time and chemoradiation therapy, this region appears to have resolved to necrotic or fluid-filled tissue zones (high $D_{ec}$, low $f_{cell}$, hypointense in FLAIR and T1w images). We note that the tumor region posterior to this cavity maintained FLAIR hyperintensity at month 8, but showed altered microstructural expression compared to the baseline month 0 scan: $f_{cell}$ is notably increased and $D_{ec}$ shows less diffuse elevation, suggesting that FLAIR hyperintensity could reflect gliotic or treatment-injured parenchyma instead of edema.[85]

Parameter maps for $f_{cell}$, $D_{ec}$, and $\kappa$ using the truncated 8-shell protocol showed alignment with fit results using the full 12-shell protocol (Figure 9), consistent with performance expected from the emerging generation of high-performance FDA approved scanners. All maps underwent degradation for the single Δ, 3-shell protocol, suggesting that TRACED-X as formulated is likely to be most reliable using a protocol compatible with these high-performance gradient scanners.

**Discussion**

This work introduces TRACED-X as a unified framework for jointly characterizing cellular microstructure, intrinsic extracellular diffusivity and tortuosity, and cell membrane permeability in solid tumors. By harnessing Monte Carlo-generated signal dictionaries, TRACED-X explicitly incorporates ECS diffusion time dependence spanning the intermediate- and long-time regimes in variable microstructures and moves beyond the simplified mono-exponential description of extracellular signal decay. This enables extracellular diffusivity to be interpreted as a biologically meaningful parameter, rather than a relative or poorly-defined quantity. Furthermore, TRACED-X incorporates water exchange beyond the Kärger model, improving estimates of microstructure and membrane permeability (Figure 2). Broadly, TRACED-X demonstrates that complex microstructural dependencies that lack analytical solutions can be effectively recovered via Monte Carlo dictionary signal inversion training, enabling new classes of parameters to be estimated in vivo in tumors.

Held-out simulation inversion tests and reduced shell in vivo fitting (Figures 1 and 9) demonstrate that the intrinsic diffusivity $D_{ec}$ was the best-recovered model parameter in TRACED-X. The fitted $D_{ec}$ may open new frontiers as a unique biomarker for characterizing the extracellular space. In vivo $D_{ec}$ maps show significant spatial heterogeneity in the intrinsic ECS diffusivity (Figure 3 and Figure 5). While the present study cannot confirm the exact biological source of $D_{ec}$ heterogeneity, it may indicate an "effective" intrinsic diffusivity after averaging (diffusion coarse-graining[33,60,71,72]) over finer-scale ECM structure in tumors, such as hyaluronic acid or collagen organization, while removing the effects of larger modeled cell boundaries. Recent work describes high-grade glioma ECM as a critical target for therapeutic development.[86] As such, we regard its observed heterogeneity as motivation for further biological validation. The estimation of $D_{ec}$ also enables a model-based calculation of ECS tortuosity via $D_{ec}/D(\Delta \rightarrow \infty)$, which may inform ECS transport restriction and drug transport models in the future.[5–7,86,87] The Monte Carlo-based formalism can be extended to any in silico architecture that can be appropriately parametrized, inviting the estimation of $\tau$ in more complex environments beyond packed spheres.

TRACED-X intracellular volume fraction estimates (Figure 6) for high grade gliomas were generally higher than previously reported values,[53,88] especially for GBM; for instance, using the VERDICT model, Roberts et al. recently estimated values in the range of ~ 0.01 - 0.135 throughout the tumor volume.[88] Our estimates of $\bar{r}$ were generally lower compared to previous estimates in GBM.[53] The directional offset (lower $f_{cell}$, elevated $\bar{r}$) in multiple prior studies is consistent with the parameter trade-off reported by Gardier et al.[27] that can occur when fitting exchange-contaminated signals without modeling the effects of exchange. We reiterate that $f_{cell}$ is a modeled representation of intracellular volume fraction, and is thus related, but not equivalent to, cell density per voxel. Therefore, the low $f_{cell}$ values demonstrated in different tumor regions may indicate that a larger volume fraction in these voxels is dominated by edema or other substrates with apparently unrestricted diffusion, while not necessarily precluding relative hypercellularity or tumor cell infiltration in these regions. TRACED-X median derived exchange times were similar to reported results using RedEx,[41] while heterogenous pockets of low $\kappa$ resulted in an increased mean and range of exchange times.

In this small cohort, multiple radiological findings in clinical imaging demonstrate coherence with TRACED-X parameter fits. In Patient 1, increased $f_{cell}$ and $\kappa$ voxels 1 colocalized with suspected mid-stage hemorrhage/clotting. Spatial organization of $f_{cell}$ and $\kappa$ values in Patient 5 (GBM) and the accompanied K-means analysis are consistent with previous work describing GBM compartmental organization (Figure 4 and Figure 7).[75] The 91.2% H4+H5 enrichment of the contrast enhancing region also agrees with recent findings from Park et al, who found that the cellular tumor (CT) core correlated with contrast enhancing regions, while the infiltrating tumor (IT) was associated with nonenhancement but elevated cellularity.[89] Our results suggest that unique tumor habitats may be identifiable from TRACED-X estimates of cellular microstructure, ECS diffusivity, and membrane permeability alone.

Clinically, the validation of such tumor subregions could provide an actionable imaging roadmap for guiding therapeutic intervention.[90] For instance, in GBM, consensus groups continue to struggle to identify the appropriate tumor area for resection,[90,91] leading to mixed results regarding its benefit. The high-to-moderately high $f_{cell}$ regions H3-H5 encompass but extend beyond the CE region – precisely underpinning the conceptual motivation for supramaximal tumor resection beyond contrast enhancement. Thus, if validated, TRACED-X maps and cluster habitats could provide improved guidance for resection or treatment of the infiltrating tumor zone.

*Limitations and Future Work*

A central limitation in diffusion MRI biophysical modeling is the reduction of complex, heterogenous tissue microstructure to a simplified representation of the intra- and extracellular spaces. Future work could incorporate more advanced microstructure modeling into the Monte Carlo signal dictionary using emergent tools.[46,92–94] For example, GBM cells at the infiltrative edge often assume a stellate shape with long processes.[95,96] Cells may align along the vasculature or white matter tracts as a scaffold for invasion, producing irregular zones of anisotropy inside the voxel.[96] Advanced acquisition strategies such as double diffusion encoding[97] could provide complimentary information regarding microscale anisotropy. Flexible intracellular diffusivity[55] or variable compartment-T2 methods (i.e., diffusion relaxometry)[98–100] would add additional levels of realism. The effects of different ECS compartments, such as edema and necrosis, were not separately modeled. Thus, we emphasize that the fitted parameters are best interpreted as effective representations of a much more complex mixture of biological contributors. The reported parameters may be interpreted as biological signatures of intra/extracellular-like heterogeneity in the tumor microenvironment to motivate additional validation (e.g., using use of ex vivo histology and/or intravital microscopy).

Figure 9 suggests that TRACED-X fitting is most stable with high-performance gradient scanners. Our preliminary results suggest that the gradient performance regime required aligns with the recent FDA approved class of scanners such as the Cima.X and MAGNUS, making such acquisitions more feasible in a growing number of clinical practices. Additional parameter constraints may increase fit stability with a more clinically-available scanners.[101]

Finally, TRACED-X may help address relevant clinical questions in tumors other than primary gliomas, including other intraaxial and extraaxial primary brain tumors and tumor metastasis. Our multiparameter estimates could also be integrated with other clinical and histopathologically derived data (including the growing number of digital pathologic features) in forming advanced multimodal AI models designed to predict prognosis and treatment response. The new generation of high-performance gradients in whole body systems also invites the application of these tools to tumors outside the brain, including prostate and breast, where diffusion imaging has already demonstrated clinical value, but where additional questions around tumor aggressiveness and treatment response remain.

## Conclusion

TRACED-X expands upon previous work by explicitly handling time-dependent, non-Gaussian extracellular diffusivity and cell membrane water permeability in complex microstructures. Proof-of-concept parameter fitting showed encouraging agreement with phenotypical expectations across high and low-grade glioma types. Future work will assess the clinical utility of the parameters we can now estimate, such as intrinsic extracellular diffusivity, tortuosity, cell size distribution, and membrane permeability, in the diagnosis and treatment of broader glioma patient populations.

**Acknowledgements:** The authors gratefully acknowledge the support of the patients and families who contributed to this research. This research was supported by the National Institutes of Health (NIH) under award numbers 5T32EB001680, P41EB030006, U01EB026996, U24NS137077, R01NS118187, R21AG085795, and DP5OD031854, as well as the MIT School of Engineering MathWorks Fellowship. This project utilized computational resources at the Martinos-MLSC cluster at the MGH Martinos Center for Biomedical Imaging Core Facility.

**Table 1.** Patient list, fitted parameters, and tumor volume.

| Patient # | Pathology Diagnosis | Approx. Tumor Volume (cc) |
|---|---|---|
| 1 | Grade 3 astrocytoma | 2.20 |
| 2 | Grade 2 astrocytoma | 0.20 |
| 3 | Grade 2 oligodendroglioma | 0.24 |
| 4 | Grade 2 astrocytoma | 0.09 |
| 5 | Glioblastoma | 1.54 |
| 6 | TG* | 0.04 |
| 7 | Grade 3 astrocytoma | 0.32 |

| 8 | LGGNT | 0.06 |
|---|---|---|

*TG, low grade tectal gioma (*no tissue, presumed); AA; LGGNT, low-grade glioneuronal tumor.*

**FIGURES:**

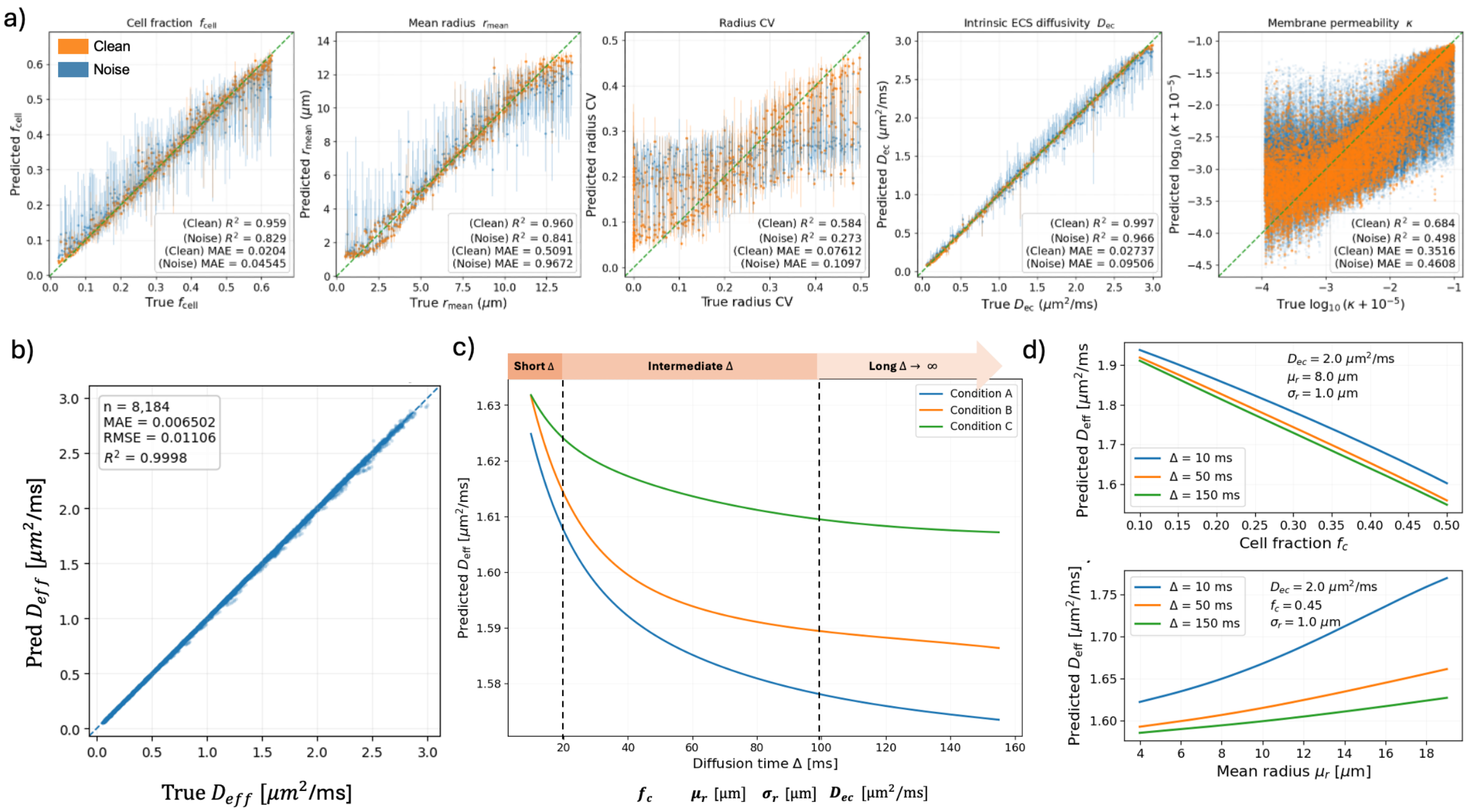

| | $f_c$ | $\mu_r$ [μm] | $\sigma_r$ [μm] | $D_{ec}$ [μm²/ms] |
|---|---|---|---|---|
| Condition A: | 0.30 | 8.0 | 1.0 | 1.82 |
| Condition B: | 0.45 | 6.0 | 1.0 | 2.00 |
| Condition C: | 0.10 | 5.0 | 3.0 | 1.68 |

**Figure 1.** a) Performance of the MLP regressors per parameter using held-out Monte Carlo test samples excluded from regressor training. Splitting was performed per geometry. For parameters besides $\kappa$, scatterplot dots represent the mean performance per geometry across all sampled $\kappa$ values, while lines show the range of predictions across $\kappa$ levels. Inline metrics are inclusive of all $\kappa$ levels across geometries. Permeability errors are reported in log-transformed space. b) $D_{eff}$ prediction vs. truth for held-out Monte Carlo test samples excluded from training. NN performance showed strong accuracy in predicting $D_{eff} = NN(D_{ec}, \mu_r, \sigma_r, f_c, \Delta)$. Simulated signals with < 5 positive values points were not considered in the analysis (8 rejected signals). c,d) Simulations demonstrating the time-dependence of $D_{eff}$ and its relationship with tissue parameters at clinically relevant diffusion times. Under three simulated microstructure conditions (A, B, C), $D_{eff}$ shows stronger time-dependence at the short Δ than intermediate $\Delta$ regimes. All curves approach an asymptote for $\Delta > 100$ ms as predicted by the diffusion coarse-graining effect.[33,60,71,72] $D_{eff}$ increases with decreasing $f_c$ and increasing $\mu_r$.

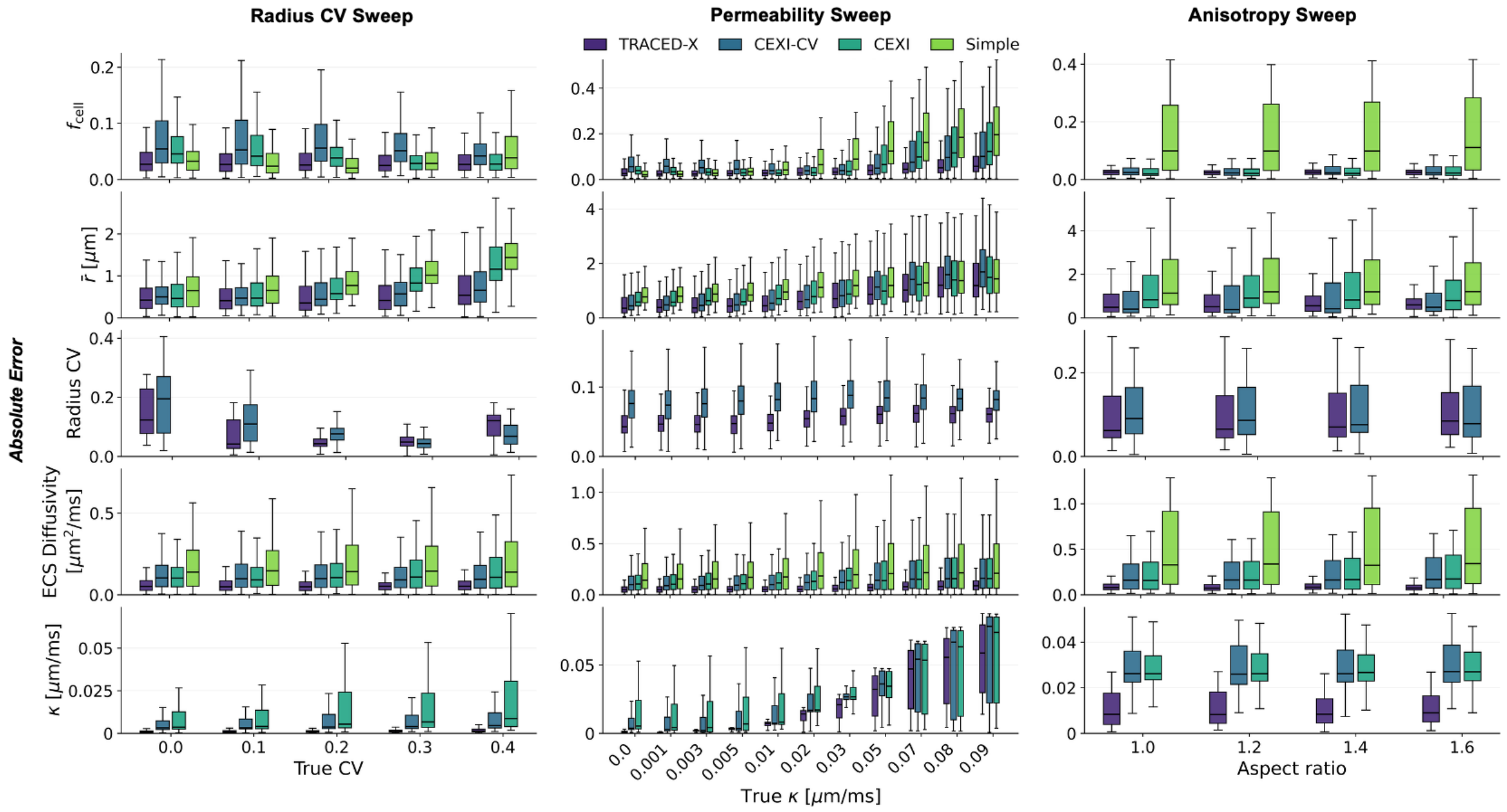

**Figure 2.** Simulation benchmarking results per parameter for TRACED-X and three comparator models (CEXI, CEXI-CV, and the Simple 2-compartment model). (Black line – median absolute error; box edges – interquartile range (IQR); whiskers – lowest (highest) observed value within Q1-1.5IQR (Q3+1.5IQR). Outliers were included in statistical analysis but not displayed). Fit performance for each parameter was assessed for held-out Monte Carlo-generated diffusion signals over a variety of microstructural conditions (radius CV, $\kappa$, and anisotropy sweeps). In most settings, TRACED-X resulted in reduced median absolute error and interquartile range compared to other models.

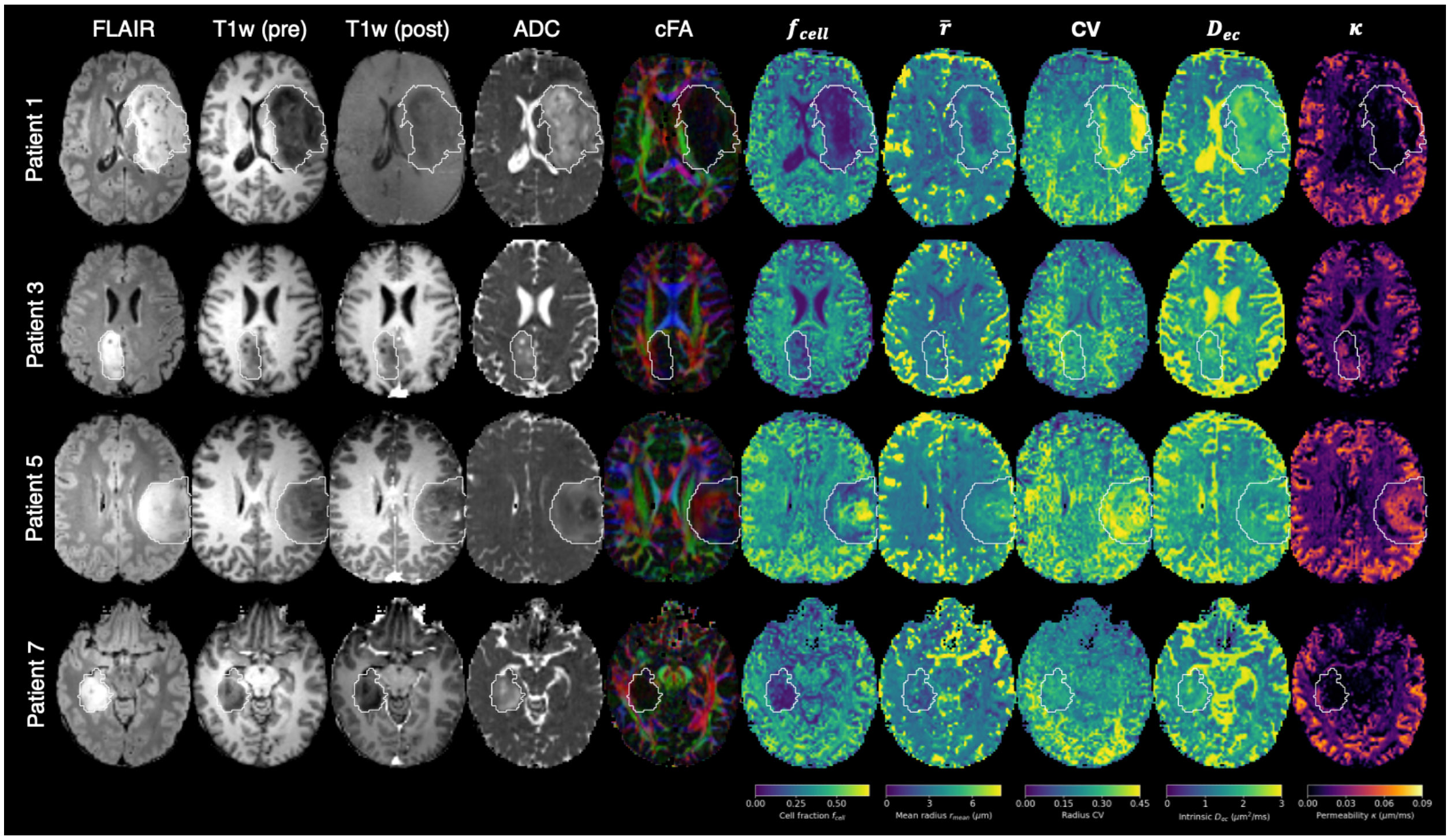

**Figure 3.** TRACED-X fitted maps for 4 patients are displayed alongside anatomical contrasts, apparent diffusion coefficient (ADC) maps, and direction- encoded color fractional anisotropy (cFA) maps. All maps display unique quantifiable contrast compared to conventional images. Patient 1 demonstrated elevated $f_{cell}$, $\bar{r}$, and radius CV near the tumor periphery with a central line of elevated $\kappa$. In Patients 3 and 7, FLAIR/T1w-images suggest tumor microcystic lesions, matched with elevated $D_{ec}$. Patient 5 showed contrast enhancement colocalized with unique microstructural and exchange behavior near the tumor core.

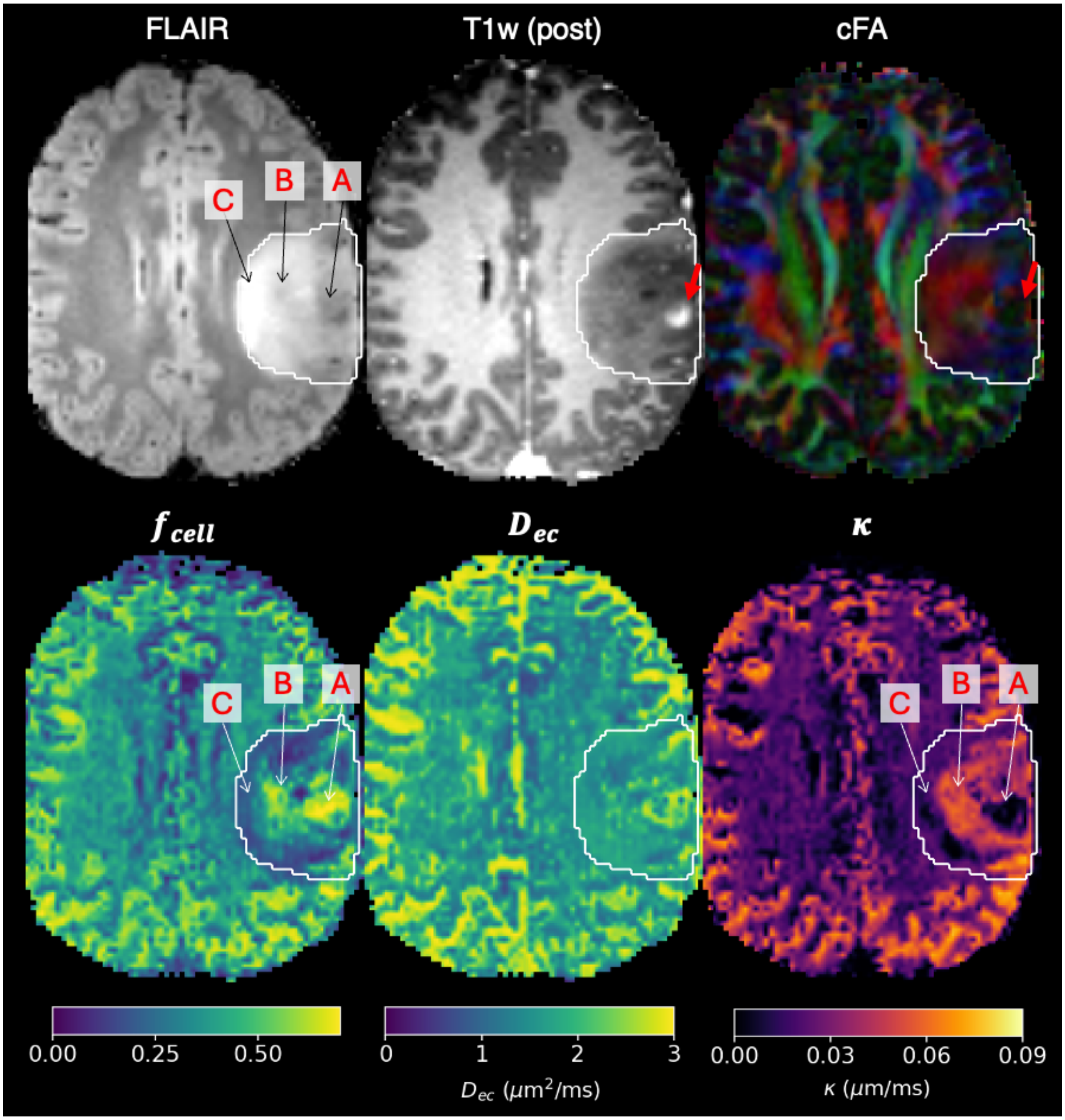

**Figure 4.** Patient 5 images showed contrast enhancement (red arrows) inside broader regions of elevated $f_{cell}$ and $D_{ec}$. Three distinct zones (A, B, C) are clearly visualized in $f_{cell}$ and $\kappa$ maps, which may correspond to the central cellular tumor (CT), infiltrating tumor (IT), and leading edge (LE).[75,89] A similar gradient of increasing hyperintensity is observed in the FLAIR image from Zone A to C. Thus, the reduced $f_{cell}$ in Zone C may be an indication that per-voxel volume fraction is increasingly dominated by edema. A reduction in fractional anisotropy (FA) is observed in the contrast-enhancing/cellular tumor region.

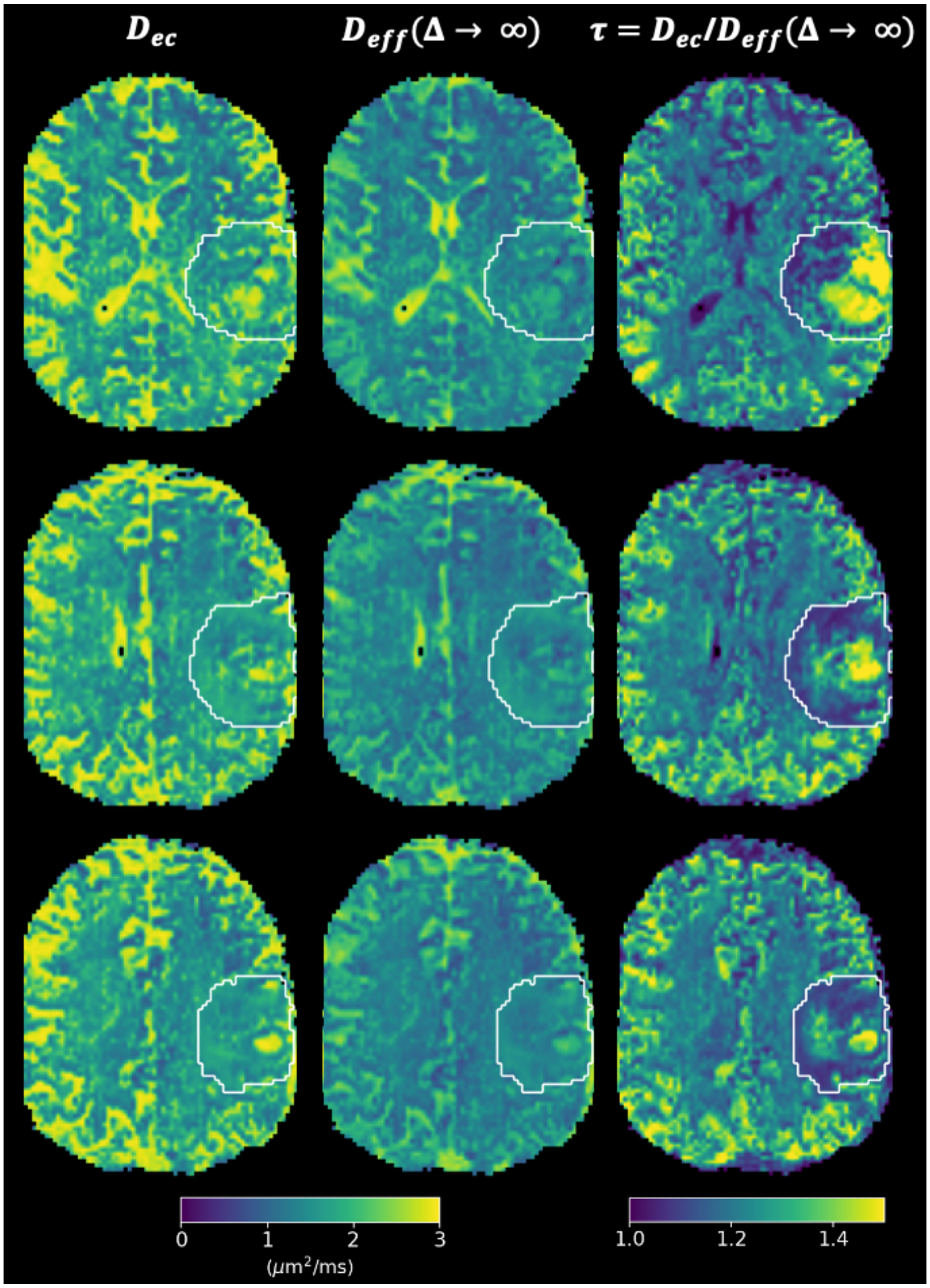

**Figure 5.** Displayed in 3 slices from Patient 5, TRACED-X natively estimates $D_{ec}$, while the surrogate neural network (see Figure 1) calculates the corresponding $D_{eff}$, which is a function of both diffusion time and microstructure. The resulting tortuosity maps ($D_{ec}/D_{eff}(\Delta \rightarrow \infty)$) represent a geometric ECS hindrance factor.

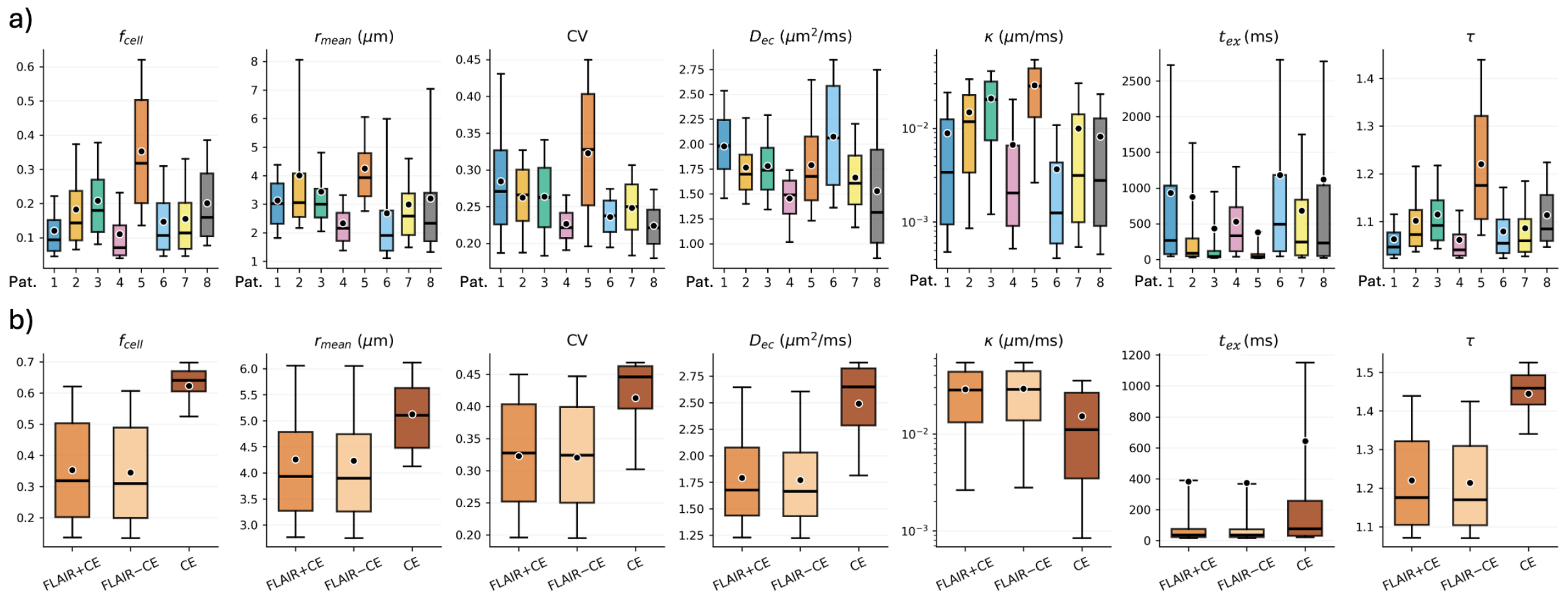

**Figure 6.** a) Median (black line), mean (black dot), IQR (boxes) and observed range within the quartile ±1.5IQR (whiskers) for fitted and derived TRACED-X parameters inside the FLAIR-hyperintense ROI demonstrate heterogeneity in all 8 mixed-grade glioma patients. Patient 5 (GBM) shows notable differences compared to lower grade gliomas. b) Whole tumor (the union of FLAIR and contrast enhancing (CE) masks - FLAIR+CE), FLAIR hyperintense only (CE tissue excluded - FLAIR-CE), and CE only ROI fit results for Patient 5. No necrosis was noted via imaging. The CE region, on average, demonstrated elevated $f_{cell}$, $\bar{r}$, $CV$, $D_{ec}$, $\tau$, $t_{ex}$, and mildly reduced $\kappa$, though habitat results (Figure 7) reveal additional heterogeneity. The derived exchange time and ECS tortuosity were also elevated in the CE region.

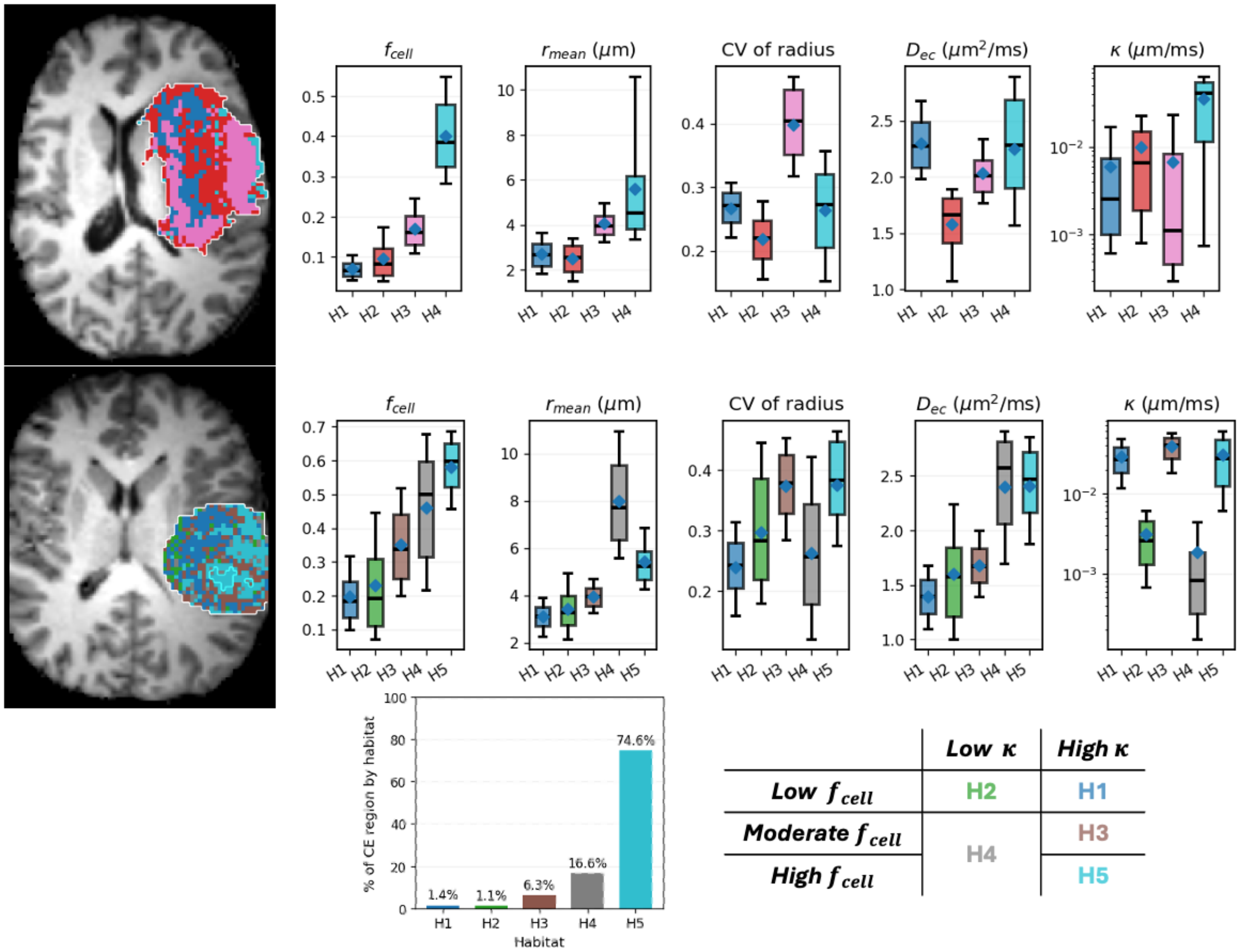

**Figure 7.** K-Means cluster analysis for Patient 1 (Grade 3 Astrocytoma) and Patient 5 (GBM). Patient 1 exhibited 4-style clusters or habitats, with moderate-to-high $f_{cell}$ habitats (H3-H4) generally localized near the tumor periphery and reduced $f_{cell}$, mixed $D_{ec}$ habitats identified predominantly in the central tumor region. Patient 5 exhibited 5 stable habitats, with unique $f_{cell} \times \kappa$ stratification. H4-H5 predominated in the central tumor region, followed by H3, and finally the lowest $f_{cell}$ H1-H2 in the tumor periphery. The contrast enhancing (CE) region was predominantly described by H4 (16.6%) and H5 (74.6%).

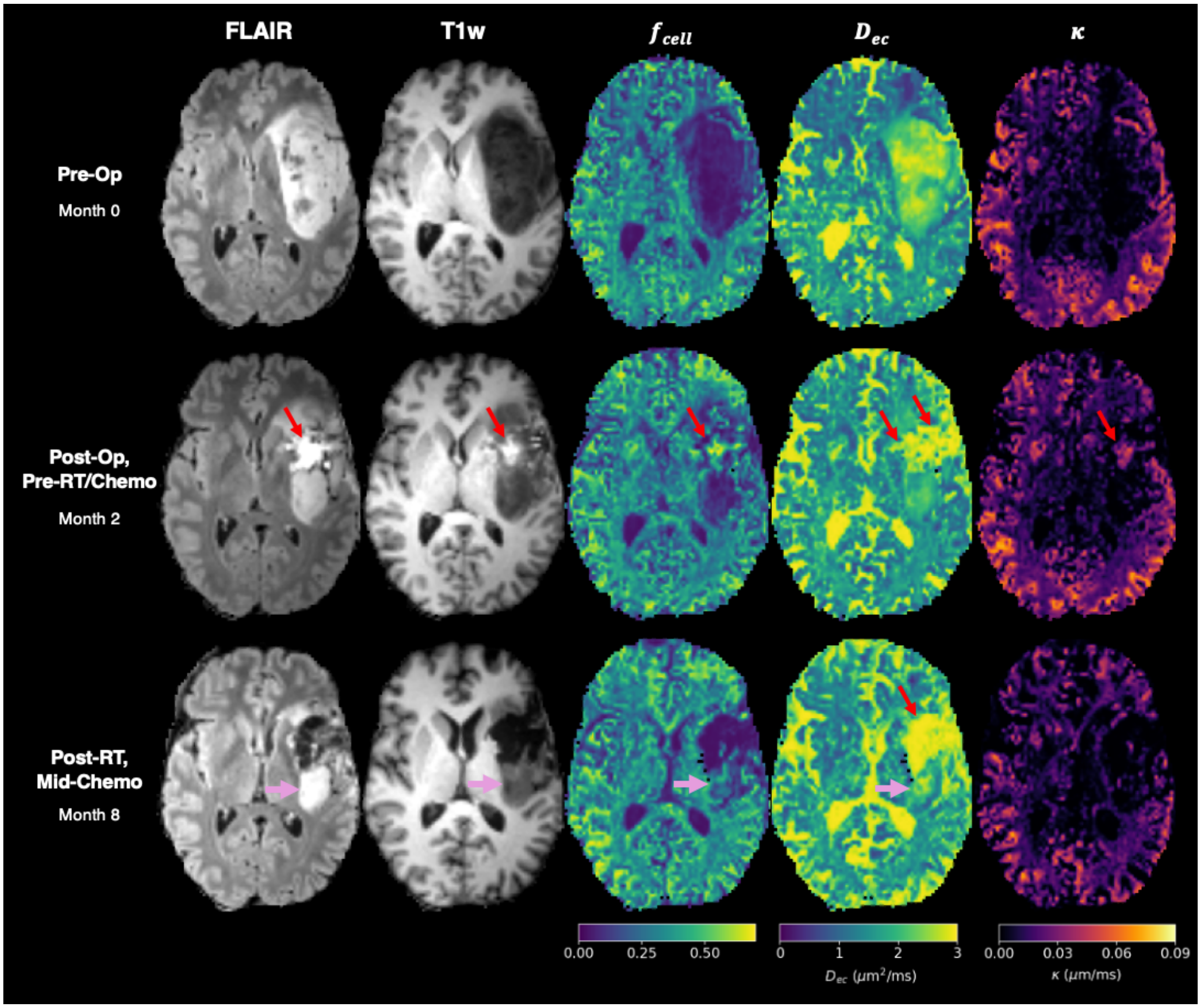

**Figure 8.** Longitudinal imaging of Patient 1. Follow-up imaging 27 days post-surgery indicated the transient appearance of suspected methemoglobin, which may have developed following delayed intratumoral hemorrhage. The T1 pre-contrast hyperintensity pattern is matched with elevated $f_{cell}$ and $\kappa$ and reduced $CV$, which may indicate the presence of clumped erythrocytes (red arrows). The same region later developed low $f_{cell}$, and elevated $D_{ec}$ (month 8) which may indicate a fluid-rich cavity in response to treatments. The tumor region posterior to the suspected cavity demonstrates FLAIR hyperintensity and T1w hypointensity similar to the baseline scan, while showing a heterogeneous increase in $f_{cell}$ (restricted volume fraction) and reduction in diffuse $D_{ec}$ elevation, suggestive of treatment related tissue changes not detected by the FLAIR or T1w images.

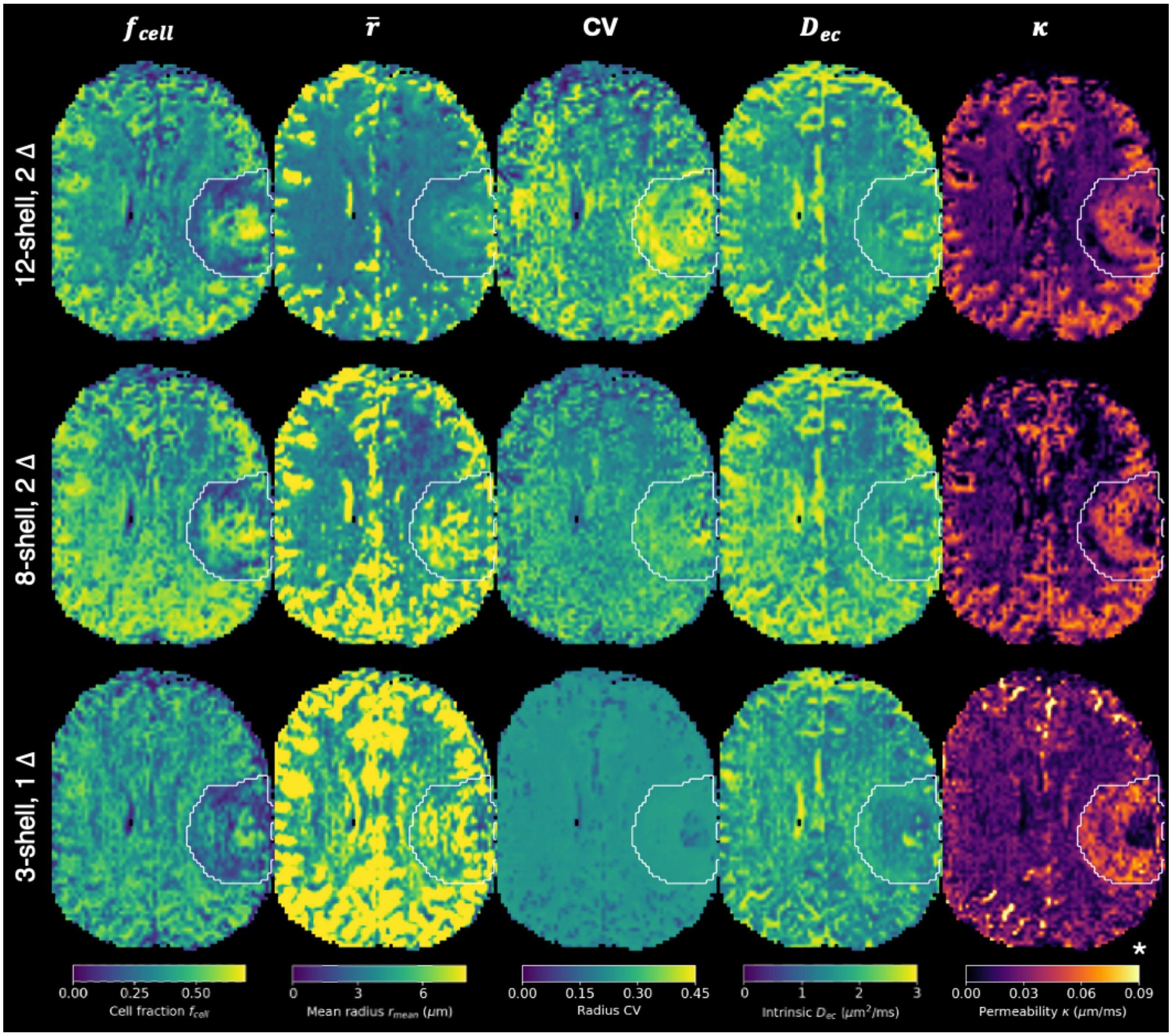

**Figure 9**. Fit performance using the full 12-shell, 2-Δ protocol, a reduced 8-shell, 2-Δ protocol, and a 3-shell, single Δ clinically-oriented protocol. Both $f_{cell}$ and $D_{ec}$ are reasonably preserved for the 8-shell protocol, while degradation is noted for the 3-shell protocol, though general organization of high/low cell fraction regions is preserved. Notable degradation of $\bar{r}$ and radius CV begins with the 8-shell protocol. The recovery of $\kappa$ is exceptionally strong for the 8-shell protocol but dramatically decreases with 3-shells. *Note that the true color scale for the 3-shell fitted $\kappa$ is 7x reduced the displayed color bar (map range adjusted for display).